\documentclass[twocolumn,pre]{revtex4}

\usepackage{dcolumn}
\usepackage{amsmath}

\usepackage{graphicx}
\usepackage{calrsfs}

\newcommand{\ii}{\mathrm{i}}
\newcommand{\half}{\tfrac12}

\newcommand{\defn}{\textit}
\newcommand{\Tr}{\mathop\mathrm{Tr}}
\renewcommand{\Im}{\mathop\mathrm{Im}}
\newcommand{\mat}{\mathbf}

\begin{document}

\title{Spectra of networks containing short loops}
\author{M. E. J. Newman}
\affiliation{Department of Physics and Center for the Study of Complex Systems,\\
University of Michigan, Ann Arbor, Michigan, USA}

\begin{abstract}
  The spectrum of the adjacency matrix plays several important roles in the mathematical theory of networks and in network data analysis, for example in percolation theory, community detection, centrality measures, and the theory of dynamical systems on networks.  A number of methods have been developed for the analytic computation of network spectra, but they typically assume that networks are locally tree-like, meaning that the local neighborhood of any node takes the form of a tree, free of short loops.  Empirically observed networks, by contrast, often have many short loops.  Here we develop an approach for calculating the spectra of networks with short loops using a message passing method.  We give example applications to some previously studied classes of networks.
\end{abstract}

\maketitle

\section{Introduction}
The adjacency matrix of an undirected network or graph with $n$ nodes is the $n\times n$ symmetric matrix~$\mat{A}$ having elements $A_{uv}=1$ if node~$u$ is connected to node~$v$ by an edge and $A_{uv}=0$ otherwise.  The spectral properties of such matrices play a central role both in the mathematical theory of networks and in practical methods for the analysis of network data.  The leading eigenvalue, for instance, is related to the percolation threshold of a network~\cite{BBCR10,KNZ14}, the leading eigenvector is widely used as a centrality measure~\cite{Bonacich87}, and spectral properties play a role in graph partitioning and community detection~\cite{Fiedler73,PSL90,Newman06b}, in localization~\cite{CM11,MZN14}, in the behavior of dynamical systems on networks~\cite{PG16}, and in detectability transitions~\cite{DKMZ11a,NN12}.

Spectra can be calculated using well established, though computationally demanding, numerical methods such as the QR algorithm, but there are also a number of analytic approaches, all closely related, that make use of message passing or cavity techniques~\cite{DGMS03,RCKT08,RPTP10,MNB11}.  In these approaches, the spectrum is expressed in terms of a set of complex-valued ``messages'' that are passed between adjacent nodes of the network, such that the values of the messages a node sends can be calculated, via fairly simple closed-form equations, from those it receives.  In practical applications the equations are usually evaluated numerically, so that these methods are not fully analytic.  Nonetheless, they can provide an estimate of the spectrum of the network that is accurate to any desired degree of precision, and they also form the foundation for a variety of additional analytic calculations, for instance of the spectra of random graphs~\cite{SC02b,Kuhn08,RCKT08,RPTP10,MNB11,NZN19}.

One disadvantage of these methods, however, is that they are in general restricted to networks that are \defn{locally tree-like}, meaning that the neighborhood of any node in the network takes the form of a tree, free of short loops, out to arbitrarily large distances in the limit of large network size.  (By ``short loops,'' we mean loops whose size remains constant as the network becomes large---a triangle would be an example of a short loop in a network.)  Unfortunately, while model networks such as random graphs are typically tree-like in this sense, real-world networks usually are not.  We would like to understand what effect the presence of short loops in networks might have on network spectra, but current analytic techniques cannot address this issue.

In this paper, we present a message passing method that allows us to compute adjacency matrix spectra for networks that do contain short loops.  Specifically, we show how to compute the spectra of networks that are made up of a collection of finite subgraphs or \defn{motifs} which may contain loops, joined together via shared nodes.  Such networks have been studied previously, for instance in the context of random graph models in which motifs are placed at random~\cite{Newman09b,Miller09,KN10b}, but as far as we are aware no calculation of their spectrum has been performed.  There are, of course, calculations going back some decades of the spectra of more specialized networks containing loops, such as regular lattices.  Husimi graphs, a generalization Bethe lattices built out of short loops of the same length, have also been studied~\cite{EKBV05,MNB11,BMN12}.  Our work, however, focuses on more general classes of networks, closer to those observed in empirical studies.

\section{Networks containing short loops}
There are a number of ways of mathematically describing, representing, or generating networks with short loops.  One common approach involves ``triadic closure,'' meaning any of several processes in which loops are added to a network that initially has none by looking for pairs of nodes with a common neighbor and connecting them to form a triangle~\cite{JGN01,HK02b,KE02,SB05}.  This approach has the advantage of mirroring directly a mechanism by which loops are believed to form in some real networks.  On the other hand, it turns out to be quite difficult to treat analytically: even describing the ensemble of networks generated by such processes is non-trivial.

An alternative approach is to create networks by directly placing loops in them.  For instance, one can create a random graph model in which not only single edges but also complete triangles are strewn among a set of nodes, each triangle connecting three randomly chosen nodes~\cite{Newman09b}.  This approach creates ensembles of networks that are relatively straightforward to describe and analyze.  One can, for instance, calculate the expected sizes of components, or percolation properties~\cite{Newman09b,Miller09,KN10b}.  One can take this approach further and create random graphs in which not merely triangles but also other, larger motifs are added to the network---sets of four, five, or more nodes, connected in any of a variety of different ways~\cite{KN10b}.

\begin{figure}
\begin{center}
\includegraphics[width=6cm]{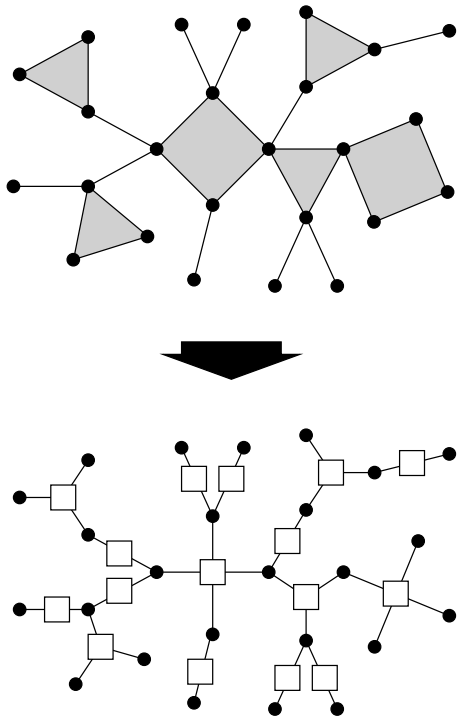}
\end{center}
\caption{Top: a network built from a collection of motifs.  In this case there are three motifs: single edges, triangles, and loops of length four.  The shading is intended only to highlight the motifs; the network itself consists of edges between nodes alone.  Bottom: the same network represented as a factor graph, a bipartite network with two sets of nodes.  One set, the filled circles, represents the nodes of the original network.  The other, the open squares, represents the motifs.}
\label{fig:factor}
\end{figure}

Here we consider networks of this latter type, but we will not focus on any particular model for generating them.  We will not, for instance, assume that they are generated randomly (although they could be).  Instead, we demand only a weaker condition on our networks, as follows.  We consider networks built of motifs, as we have described, where each motif is a connected subgraph joining together a specific set of nodes in some specific pattern of edges.  A single edge connecting two nodes is itself considered to be a motif, the simplest possible example.  A triangle is another example, and there are many larger possibilities.  Networks built in this fashion can be represented by a \defn{factor graph}, a bipartite graph having two distinct sets of nodes.  One set represents the nodes of the original network, the other represents the motifs, and there are edges connecting each node to the motifs to which it belongs---see Fig.~\ref{fig:factor}.  In our calculations we will assume that this factor graph is locally tree-like.  The original network itself is not in general tree-like---it may contain many loops of various lengths that fall within the individual motifs---but the factor graph is tree-like.  In effect, we are saying that all the short loops in the network are accounted for within the motifs.  There are no additional short loops other than these.  Our approach works by writing the spectrum of the network in terms of a message passing algorithm that acts not upon the original network but upon the factor graph, whose locally tree-like structure then makes the calculation exact, at least in the limit of large network size.

\section{Message passing}
We develop our message passing approach for the general case of networks formed of motifs of any size and structure, arranged on a locally tree-like factor graph as in Fig.~\ref{fig:factor}.  In Section~\ref{sec:triangles} we apply our approach to the specific example of a network formed of just two motifs, single edges and triangles, for which the message passing equations take a particularly simple form.

\subsection{Spectral density}
Our goal will be the calculation of the \defn{spectral density} or \defn{density of states}~$\rho(x)$ of a given network of $n$ nodes, which is the function
\begin{equation}
\rho(x) = {1\over n} \sum_{i=1}^n \delta(x-\lambda_i),
\label{eq:density}
\end{equation}
where $\lambda_i$ is the $i$th eigenvalue of the adjacency matrix and $\delta(x)$ is the Dirac delta function.  Following a standard line of argument, we write the delta function as the limit of a suitably normalized Lorentzian (or Cauchy) distribution as its width~$\eta$ tends to zero:
\begin{equation}
\delta(x) = \lim_{\eta\to0^+} {\eta/\pi\over x^2+\eta^2}
  = -{1\over\pi} \lim_{\eta\to0^+} \Im {1\over x+\ii\eta},
\label{eq:delta}
\end{equation}
where the notation $\eta\to0^+$ indicates that~$\eta$ tends to zero from above.  Substituting this form into Eq.~\eqref{eq:density} we get
\begin{equation}
\rho(x) = -{1\over n\pi} \lim_{\eta\to0^+} \Im \sum_{i=1}^n
           {1\over x-\lambda_i+\ii\eta}.
\end{equation}
It will be convenient to define a generalization of the spectral density to the complex plane thus:
\begin{equation}
\rho(z) = -{1\over n\pi} \sum_{i=1}^n {1\over z-\lambda_i}
        = -{1\over n\pi} \Tr (z\mat{I}-\mat{A})^{-1}.
\label{eq:complexrho}
\end{equation}
The standard real spectral density~$\rho(x)$ is then given by the imaginary part of this quantity where $z = x+\ii\eta$ and we take the limit as $\eta$ goes to zero from above.  The quantity~$\eta$ acts as a broadening parameter that broadens the delta-function peaks in the spectrum by an amount roughly equal to~$\eta$, and in practical calculations on finite networks it is sometimes convenient to retain a small non-zero value of $\eta$ in order to make~$\rho(x)$ a smooth function of its argument.

Expanding the matrix inverse in Eq.~\eqref{eq:complexrho} as a geometric series $(z\mat{I}-\mat{A})^{-1} = z^{-1} \sum_{r=0}^\infty (\mat{A}/z)^r$ and taking the trace term by term, we have
\begin{equation}
\rho(z) = -{1\over n\pi z} \sum_{r=0}^\infty {\Tr \mat{A}^r\over z^r}.
\label{eq:rhotrace}
\end{equation}
The quantity $\Tr\mat{A}^r$ is equal to the number of closed walks of length~$r$ on the network, a closed walk being any (possibly self-intersecting) path across the network that starts and ends at the same node.  If we can count the number of such walks for all lengths~$r$ then we can calculate the spectral density from Eq.~\eqref{eq:rhotrace}.

\subsection{Counting closed walks}
Let $u$ denote an arbitrary node in the network and $\sigma$ denote one of the motifs to which it belongs.  Because the factor graph is locally tree-like, any closed walk that starts at node~$u$ and takes its first step across one of the edges in motif~$\sigma$ must, on its final step, return to node~$u$ also across one of the edges in~$\sigma$ (although not necessarily the same edge).  Were this not the case, were the walk to return via a different motif, then it would in the process complete a loop on the factor graph, of which, by hypothesis, there are none, and hence such walks cannot exist.

Let us define $N^{u\sigma}_r$ to be the number of closed walks that start at node~$u$, take their first step along one of the edges in~$\sigma$, and return to node~$u$ for the first time, also via~$\sigma$, exactly $r$ steps later.  Any node other than~$u$ may be visited as many times as we wish during the walk, but node~$u$ is visited only twice, once at the start of the walk and once at the end.  We will call such walks \defn{excursions}.

The total number of distinct possible excursions of length~$r$ from node~$u$ is given by the sum of $N^{u\sigma}_r$ over the set~$\mathcal{S}_u$ of all motifs~$\sigma$ to which $u$ belongs: $\sum_{\sigma\in\mathcal{S}_u} N^{u\sigma}_r$.  Walks of length~$r$ that visit their starting node more than twice---say $m$ times other than at the start of the walk---are made up of $m$ distinct excursions with lengths~$r_1\ldots r_m$ such that $\sum_{i=1}^m r_i = r$.  Thus the number~$N^u_{rm}$ of such walks is
\begin{equation}
N^u_{rm} = \sum_{r_1=1}^\infty \ldots \sum_{r_m=1}^\infty
  \delta\bigl( r, {\textstyle \sum_i r_i} \bigr)\prod_{i=1}^m\>
  \sum_{\sigma\in\mathcal{S}_u} N^{u\sigma}_{r_i},
\label{eq:nrm}
\end{equation}
where $\delta(i,j)$ is the Kronecker delta.  (Terms with $r_i=1$ can only appear in networks that have self-loops---edges that connect a node to itself---which is rare in real-world situations.  In all other networks the shortest possible excursion has length~2.  We leave these terms in the expression for the sake of completeness, however.)

Summing Eq.~\eqref{eq:nrm} over all possible values of $m$ and all nodes~$u$ now gives us the total number of closed walks of length~$r$ in the whole network with any number of excursions, which is precisely equal to the quantity~$\Tr \mat{A}^r$ that we are trying to calculate.  Substituting into Eq.~\eqref{eq:rhotrace} we then get
\begin{align}
\rho(z) &= - {1\over n\pi z} \sum_{r=0}^\infty {1\over z^r}
           \sum_{u=1}^n\>\sum_{m=0}^\infty N^u_{rm} \nonumber\\
        &= - {1\over n\pi z} \sum_{r=0}^\infty {1\over z^r}
           \sum_{u=1}^n\>\sum_{m=0}^\infty\>
           \sum_{r_1=1}^\infty \ldots \sum_{r_m=1}^\infty
           \delta\bigl( r, {\textstyle \sum_i r_i} \bigr) \nonumber\\
        &\hspace{16em}{} \prod_{i=1}^m\>
           \sum_{\sigma\in\mathcal{S}_u} N^{u\sigma}_{r_i} \nonumber\\
        &= - {1\over n\pi z} \sum_{u=1}^n
            \sum_{m=0}^\infty\>\prod_{i=1}^m\>
            \sum_{r_i=1}^\infty\>\sum_{\sigma\in\mathcal{S}_u} 
            {N^{u\sigma}_{r_i}\over z^{r_i}},
\label{eq:rhomp1}
\end{align}
where we adopt the convention that the empty product $\prod_{i=1}^0$ is equal to~1.  Defining the useful quantity
\begin{equation}
\mu_{u\sigma}(z) = \sum_{r=1}^\infty {N^{u\sigma}_r\over z^{r-1}},
\label{eq:defsmu}
\end{equation}
we can write Eq.~\eqref{eq:rhomp1} as
\begin{align}
\rho(z) &= - {1\over n\pi z} \sum_{u=1}^n\>\sum_{m=0}^\infty
            \biggl[ \sum_{\sigma\in\mathcal{S}_u} {\mu_{u\sigma}(z)\over z}
            \biggr]^m \nonumber\\
        &= - {1\over n\pi} \sum_{u=1}^n
           {1\over z - \sum_{\sigma\in\mathcal{S}_u} \mu_{u\sigma}(z)}.
\label{eq:rhosoln}
\end{align}

We regard $\mu_{u\sigma}(z)$ as a ``message,'' sent by motif~$\sigma$ to node~$u$, whose value can be calculated as we now demonstrate.

\subsection{Message passing equations}
\begin{figure}
\begin{center}
\includegraphics[width=6cm]{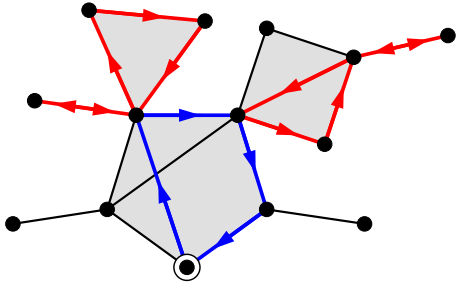}
\end{center}
\caption{A complete closed walk from a starting node (bottom, circled) consists of a closed walk within a single motif, here consisting of five nodes, plus, optionally, any number of excursions along the way that leave the motif and return to it some time later via the same node.  Each excursion is itself a closed walk of the same type, which allows us to write a self-consistent expression~\eqref{eq:nwalks} for the number of walks.}
\label{fig:walk}
\end{figure}

To evaluate the messages~\eqref{eq:defsmu} we need to compute the number of excursions~$N^{u\sigma}_r$---the number of walks from $u$ that take their first (and last) step via motif~$\sigma$ and visit $u$ only at the start and end of the walk.  The structure of such walks is illustrated in Fig.~\ref{fig:walk}.  Each consists of a closed walk around motif~$\sigma$ itself, which visits $u$ only at its start and end, plus any number of excursions from nodes in~$\sigma$ (other than~$u$) to the rest of the network.

Let $w$ be a walk of length~$k+1$ around motif~$\sigma$, which we will represent by the list $v_1\ldots v_k$ of $k$ nodes (not necessarily distinct) that the walk passes through other than node~$u$, and let $r^v_1, r^v_2,\ldots$ be the lengths of the excursions from node~$v$.  Then a complete walk of length~$r$, excursions included, must have $r = k + 1 + \sum_{v\in w} \sum_i r^v_i$, and the total number of complete walks of length~$r$ that have $w$ as their foundation is given by
\begin{align}
&\biggl[ \sum_{m_{v_1}=0}^\infty\>\sum_{r^{v_1}_1=1}^\infty \ldots\!\!\!
  \sum_{r^{v_1}_{m_{v_1}}=1}^\infty \biggr] \ldots
\biggl[ \sum_{m_{v_k}=0}^\infty\>\sum_{r^{v_k}_1=1}^\infty \ldots\!\!\!
  \sum_{r^{v_k}_{m_{v_k}}=1}^\infty \biggr] \nonumber\\
&\hspace{2em} \delta\bigl( r, {\textstyle k + 1 + \sum_{v\in w} \sum_{i=1}^{m_v}
  r^v_i} \bigr)\> \prod_{v\in w}\>\prod_{i=1}^{m_v}\>
  \sum_{\substack{\tau\in\mathcal{S}_v\\ \tau\ne\sigma}} N^{v\tau}_{r^v_i},
\label{eq:nwalks}
\end{align}
where $m_{v_1}\ldots m_{v_k}$ represent the numbers of excursions from each of the nodes~$v_1\ldots v_k$.

The total number~$N^{u\sigma}_r$ of paths is now equal to the sum of Eq.~\eqref{eq:nwalks} first over the complete set~$\mathcal{W}^{k}_{u\sigma}$ of walks of length~$k+1$ in motif~$\sigma$ that start and end at node~$u$, and second over all $k=0\ldots\infty$.  Taking the resulting expression and substituting it into Eq.~\eqref{eq:defsmu}, we find that
\begin{align}
\mu_{u\sigma}(z) &= \sum_{k=0}^\infty
  {1\over z^k}\!\sum_{w\in\mathcal{W}^k_{u\sigma}}\>
  \prod_{v\in w}\>\sum_{m=0}^\infty\>
  \prod_{i=1}^m\>\sum_{r^v_i=1}^\infty\>
  \sum_{\substack{\tau\in\mathcal{S}_v\\ \tau\ne\sigma}}
  {N^{v\tau}_{r^v_i}\over z^{r^v_i}} \nonumber\\
  &= \sum_{k=0}^\infty {1\over z^k}\!\sum_{w\in\mathcal{W}^k_{u\sigma}}\>
     \prod_{v\in w}\>\sum_{m=0}^\infty
     \biggl[ \sum_{\substack{\tau\in\mathcal{S}_v\\ \tau\ne\sigma}}\,
     \sum_{r=1}^\infty {N^{v\tau}_r\over z^r} \biggr]^m \nonumber\\
  &= \sum_{k=0}^\infty\>\sum_{w\in\mathcal{W}^k_{u\sigma}}\>
     \prod_{v\in w} {1\over z -
     \sum_{\substack{\tau\in\mathcal{S}_v\\ \tau\ne\sigma}} \mu_{v\tau}(z)}.
\label{eq:mpderiv}
\end{align}

Defining a new message~$g_{\sigma u}(z)$, passed from node~$u$ to motif~$\sigma$, by
\begin{equation}
g_{\sigma u}(z) = {1\over z -
  \sum_{\substack{\tau\in\mathcal{S}_u\\ \tau\ne\sigma}} \mu_{u\tau}(z)},
\label{eq:mp1}
\end{equation}
we can write Eq.~\eqref{eq:mpderiv} as
\begin{equation}
\mu_{u\sigma}(z) =  \sum_{k=0}^\infty\> \sum_{w\in\mathcal{W}^k_{u\sigma}}\,
     \prod_{v\in w} g_{\sigma v}(z).
\label{eq:mp2}
\end{equation}

Equations~\eqref{eq:mp1} and~\eqref{eq:mp2} give us a complete set of self-consistent equations whose solutions give the values of the messages~$\mu_{u\sigma}(z)$.  For any given network, we can solve these equations, for example by simple iteration, to find the values of the~$\mu_{u\sigma}(z)$ for any~$z$ and then substitute the results into Eq.~\eqref{eq:rhosoln} to get the spectral density.

\section{Examples}
\label{sec:triangles}
In practice the message passing equations \eqref{eq:mp1} and~\eqref{eq:mp2} can be difficult to solve because they require us to enumerate all closed walks of a given length for every motif in the network, which in many cases is not an easy task.  One important case that is relatively straightforward, however, is the case of a network composed solely of single edges and triangles, which has been studied in some detail in the past~\cite{Newman09b,Miller09,KN10b}.

For networks of this kind, which we will call \defn{edge--triangle networks}, there are only two distinct motifs: single edges, which connect two nodes to one another, and triangles, which connect three.  Let us treat these in turn.

The case of single edges is straightforward.  There is only one closed walk in such a motif, having two steps, along the edge and back again.  In this case Eq.~\eqref{eq:mp2} simplifies to
\begin{equation}
\mu_{u\sigma}(z) = g_{\sigma v}(z),
\label{eq:edges}
\end{equation}
where $v$ is the node at the other end of edge~$\sigma$ from~$u$.

The case of triangles is only a little more complicated.  Every closed walk from $u$ in a triangle has the same form: we walk from $u$ to one of the two other nodes in the triangle---let us call them $v$ and~$w$---then we alternate back and forth between $v$ and $w$ some number of times before returning to $u$ on the final step.  For even values of $k$ in Eq.~\eqref{eq:mp2} nodes $v$ and $w$ are visited the same number of times, namely~$\half k$ each.  For odd values, one node is visited $\half(k+1)$ times and the other $\half(k-1)$ times.  Setting $k=2l$ and summing over integer values of~$l$, Eq.~\eqref{eq:mp2} can then be written
\begin{align}
\mu_{u\sigma}(z) &= \sum_{l=1}^\infty \bigl[ 2 g_{\sigma v}^l g_{\sigma w}^l +
  g_{\sigma v}^l g_{\sigma w}^{l-1} + g_{\sigma v}^{l-1} g_{\sigma w}^l \bigr] \nonumber\\
  &= {2g_{\sigma v}(z)g_{\sigma w}(z)+g_{\sigma v}(z)+g_{\sigma w}(z)
      \over 1-g_{\sigma v}(z) g_{\sigma w}(z)}.
\label{eq:triangles}
\end{align}

Combining Eq.~\eqref{eq:edges} (for single edges) and \eqref{eq:triangles} (for triangles) with Eq.~\eqref{eq:mp1}, we now have our complete message passing equations for edge--triangle networks.

\subsection{Regular graphs}
\label{sec:regular}
As an example, consider a random network in which every node belongs to exactly one single edge and one triangle.  This is the equivalent for an edge--triangle network of the regular graphs of traditional graph theory---networks in which every node has the same degree.  The spectrum of a random regular graph obeys the well-known Kesten--McKay distribution~\cite{McKay81}.  Here we calculate the equivalent result for the edge--triangle case.

A random edge--triangle network necessarily has a locally tree-like factor graph for the same reason that traditional random graphs are locally tree-like: in a random graph the probability of creating a loop vanishes in the limit of large network size.  This means our message passing equations are applicable to the random case and will be exact in the limit of large network size.

Moreover, in a \emph{regular} edge--triangle network every node is equivalent, having the same local neighborhood, out to arbitrary distances in the limit of large network size.  Similarly, every single edge has the same neighborhood as every other, as does every triangle.  This means in practice that $\mu_{u\sigma}(z)$ takes only two values, one for single edges and one for triangles, which we will denote by $\mu(z)$ and $\nu(z)$ respectively.  Similarly, $g_{\sigma u}(z)$~takes only two values, which we will denote $g(z)$ and~$h(z)$.  In terms of these quantities, the message passing equations read
\begin{equation}
\mu(z) = g(z), \qquad
\nu(z) = {2h(z)\over 1-h(z)},
\label{eq:munu}
\end{equation}
and
\begin{equation}
g(z) = {1\over z-\nu(z)}, \qquad
h(z) = {1\over z-\mu(z)}.
\label{eq:gh}
\end{equation}
Eliminating $g$ and substituting into Eq.~\eqref{eq:rhosoln}, we find that the complex spectral density~$\rho(z)$ is given by
\begin{equation}
\rho(z) = - {1/\pi\over z - \mu(z) - \nu(z)}
        = {1/\pi\over\mu - 1/\mu}.
\label{eq:strho}
\end{equation}
At the same time, eliminating $g$, $h$, and~$\nu$ from Eqs.~\eqref{eq:munu} and~\eqref{eq:gh} yields a quadratic equation for~$\mu$ with solution
\begin{equation}
\mu(z) = {z^2-z-1 \pm \sqrt{z^4-2z^3-5z^2+6z+1}\over2z}.
\label{eq:musoln}
\end{equation}
We substitute this result into~\eqref{eq:strho} and, after some manipulation, find that
\begin{equation}
\rho(z) = {z^2-z-1\pm (2z-1)
  \sqrt{z^4 - 2z^3 - 5z^2 + 6z + 1} \over
  2\pi (z^4 - 2z^3 - 5z^2 + 6z)}.
\label{eq:rhoz}
\end{equation}
Letting $z$ go to the real line at $x$, taking the imaginary part, and noting that the discriminant within the square root can be written in the form $[(x-\half)^2-\tfrac{13}{4}]^2-8$, we then recover the real spectral density
\begin{equation}
\rho(x) = {1\over\pi}\bigl|x-\half\bigr|\,
  {\sqrt{8-\bigl[\bigl(x-\half\bigr)^2-\tfrac{13}{4}\bigr]^2} \over
  9 - \bigl[\bigl(x-\half\bigr)^2-\tfrac{13}{4}\bigr]^2}.
\label{eq:rhoreg}
\end{equation}
This is the analog of the Kesten--McKay distribution for this random regular edge--triangle network.

\begin{figure}
\begin{center}
\includegraphics[width=\columnwidth]{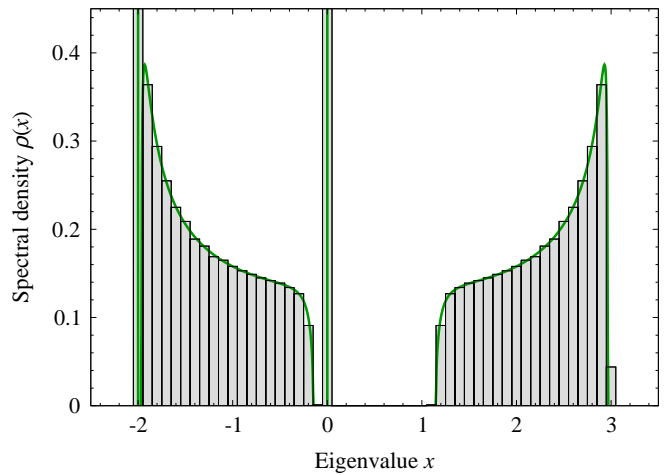}
\end{center}
\caption{The spectrum of a random regular network in which every node belongs to exactly one single edge and one triangle.  The histogram shows the distribution of eigenvalues calculated by numerical diagonalization of the adjacency matrix of one random realization of the network with $10\,002$ nodes.  The solid curves show the analytic solution.}
\label{fig:rhoreg}
\end{figure}

In order for~\eqref{eq:rhoreg} to be real, we require that the quantity within the square root be positive.  The zeros of this quantity satisfy $(x-\half)^2-\tfrac{13}{4} = \pm\sqrt{8}$, or
\begin{equation}
x  = \frac12 \biggl( 1\pm\sqrt{13\pm8\sqrt{2}} \biggr),
\end{equation}
and hence the spectral density is non-vanishing in two different bands, the first between the $x$ values $\half(1-\sqrt{13\pm8\sqrt{2}})$, which is approximately $x\in [-1.965,-0.149]$, and the second between the values $\half(1+\sqrt{13\pm8\sqrt{2}})$, which is approximately $x\in [1.149,2.965]$.  Moreover, since Eq.~\eqref{eq:rhoreg} is a function of $|x-\half|$ only, the spectrum must be symmetric about the point $x=\half$, meaning that the two bands are mirror images of one another.  Figure~\ref{fig:rhoreg} shows the shape of the spectrum computed from Eq.~\eqref{eq:rhoreg}.

Also visible in the figure are two delta-function peaks in the spectrum at $x=-2$ and $x=0$.  These appear in the solution as divergences in~$\rho(z)$ at the corresponding points, which map to delta functions via Eq.~\eqref{eq:delta}.  We could, if we wish, introduce a small Lorentzian broadening via the parameter~$\eta$ in~\eqref{eq:delta} to make these peaks visible in the figure, but we have not done that here.  Instead we have merely added the peaks to the figure by hand at their calculated positions.

We also show in Fig.~\ref{fig:rhoreg} a histogram of the spectrum, calculated by direct numerical diagonalization of the adjacency matrix of a single example network randomly generated from the model with $n=10\,002$ nodes.  (The number of nodes must be a multiple of six to satisfy the requirement that the number of ends of edges be even and the number of corners of triangles be a multiple of three.)  As the figure shows, the exact solution and numerical results agree well.

In principle, one could construct a closed-form solution analogous to~\eqref{eq:rhoreg} for the spectrum of any random regular edge--triangle network such that every node has the same number of edges and triangles.  In practice, however, the general case involves solving a quartic rather than a quadratic equation, and, while this can be done, the solution is complicated and we do not give it here.

\subsection{Random graphs}
A more complex example is a general random edge--triangle network in which we specify separately the number of edges and triangles that each node participates in.  To compute the spectral density of such networks we must solve the full message passing equations~\eqref{eq:mp1}, \eqref{eq:edges}, and~\eqref{eq:triangles}, which we do here by simple iteration.  For a given network, we choose any convenient starting values for the messages (setting them all to zero works well), then numerically iterate the equations until they converge to the desired degree of accuracy.  This process does not give us an analytic form for the spectrum, but it can in principle give us a solution accurate to any required precision.

\begin{figure}
\begin{center}
\includegraphics[width=\columnwidth]{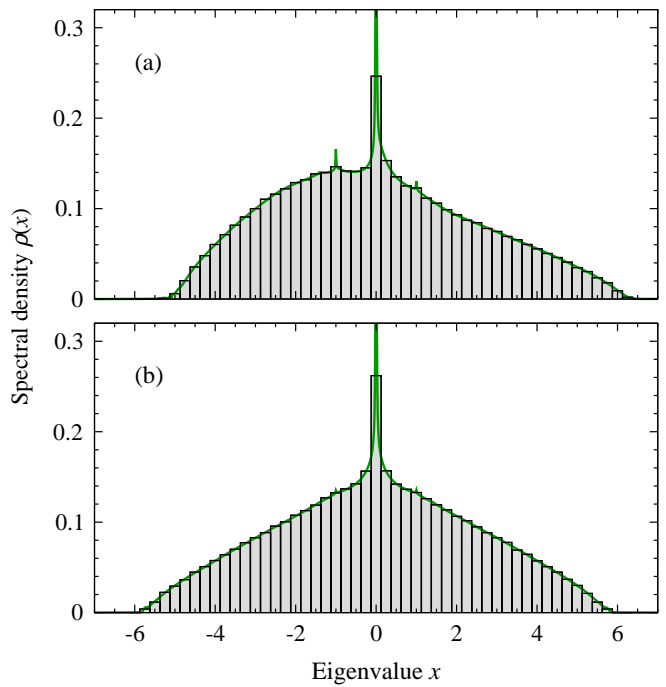}
\end{center}
\caption{(a)~The spectrum of a random network composed of single edges and triangles, with a Poisson distribution for both: the number of both edges and triangles is Poisson distributed at every node with mean~2.  (b)~The spectrum of a configuration model with the same node degrees as the network in~(a).  In each panel the solid curve shows the spectral density for a single randomly generated network with $10\,000$ nodes, calculated by direct iteration of the message passing equations with Lorentzian broadening parameter~$\eta=0.01$.  The histogram shows the spectrum of the same network calculated by numerical diagonalization of the adjacency matrix.}
\label{fig:poisson}
\end{figure}

Figure~\ref{fig:poisson}a shows the results of one such calculation.  In this example we have generated a network of $n=10\,000$ nodes, each having a number of single edges drawn from a Poisson distribution with mean~2, and a number of triangles drawn from a Poisson distribution also with mean~2.  The net result is a random network with average degree~6 and $\tfrac23 n$ triangles in total, on average.  The histogram in Fig.~\ref{fig:poisson}a shows the spectrum of the network calculated by conventional numerical diagonalization of the adjacency matrix.  The continuous curve shows the spectral density calculated using the message passing equations, with a Lorentzian broadening parameter~$\eta=0.01$ (see Eq.~\eqref{eq:delta} and the following discussion).  As the figure shows, the agreement between the two calculations is good.  Note in particular how the delta-function peaks in the spectrum at $x=0, \pm1$ emerge clearly in the message passing calculation because of the broadening.  Note also the very different overall shape of the spectrum in this case from that of Fig.~\ref{fig:rhoreg} for the random regular edge--triangle network.

How does the presence of loops in the network affect the spectrum in practice?  To shed light on this question, we show in Fig.~\ref{fig:poisson}b the spectrum of a random graph with the same node degrees as in Fig.~\ref{fig:poisson}a, but without loops.  This network is generated using the standard configuration model in which single edges only are placed between nodes in the appropriate numbers, but no triangles~\cite{MR95,NSW01}.  In the limit of large network size this produces a locally tree-like network.  Figures~\ref{fig:poisson}a and~\ref{fig:poisson}b are plotted on the same scales and, as we can see, there are some general similarities between the two but also some clear differences.  The approximate ranges spanned by the spectra are similar and both show a prominent peak at $x=0$.  But without loops the spectrum is symmetric, whereas for the network with loops it has a distinctly asymmetric shape---the presence of loops produces a clear redistribution of values, for instance shifting the upper edge of the band to a higher value.  A similar but more pronounced asymmetry is visible in the spectrum of the regular edge--triangle network in Fig.~\ref{fig:rhoreg}.

Mathematically the asymmetry of the spectrum has a straightforward explanation.  In a loopless network all closed walks have an even number of steps, and hence the series expansion in Eq.~\eqref{eq:rhotrace} contains only even powers of~$z$.  Once triangles are introduced into the network we can have walks of odd length and hence the function can contain both even and odd powers of~$z$.

The difference between the spectra for networks with and without loops indicates that calculations in which network spectral properties are approximated using model networks like the configuration model will in general produce not only quantitative but also qualitative errors because of their neglect of loops.

\section{Conclusions}
In this paper we have derived and demonstrated a message passing method for calculating the adjacency matrix spectra of networks that contain short loops.  Previous analytic approaches to calculating network spectra have typically made the assumption that the networks studied were locally tree-like, an assumption that is strongly violated by most real-world networks.  Our approach gets around this assumption by representing networks as a collection of motifs, which are allowed to contain loops, but assuming that the factor graph describing the connections between motifs is tree-like.  This assumption is enough to allow us to write down message passing equations for the spectrum of such a network, even though the network itself contains loops.

We have given two applications of our method to example networks composed of just the two simplest motifs, single edges and triangles.  In our first example we consider a random regular graph model in which each node belongs to exactly one edge and one triangle.  For this model we are able to solve the message passing equations exactly and derive a closed-form expression for the spectral density.  In our second example, we generate random networks with Poisson distributions of their numbers of edges and triangles, and compute their spectra by numerical solution of the message passing equations.  In both cases we find good agreement with direct calculations of the eigenvalues by numerical diagonalization.

The calculations described here could be extended in a number of ways.  We have considered only one example of the random regular edge--triangle network but, as mentioned in Section~\ref{sec:regular}, it would be possible in principle to calculate a closed-form expression for the spectra of such networks in the general case where nodes belong to any number of edges and triangles, although such an expression, whose derivation involves the solution of a quartic equation, seems likely to be complicated.  One could also consider more elaborate networks built of larger motifs than the simple edges and triangles we have considered here.  To do this, one would have to enumerate all closed walks within each allowed motif in order to evaluate Eq.~\eqref{eq:mp2}, which becomes progressively harder as the motifs become larger.  Some cases are relatively tractable, however.  Walks on motifs consisting of simple loops of any length, for instance, are quite straightforward to enumerate.  These issues, however, we leave for future work.

\bigskip\begin{acknowledgments}
This work was funded in part by the US National Science Foundation under grant DMS--1710848.
\end{acknowledgments}


\begin{thebibliography}{10}
\expandafter\ifx\csname url\endcsname\relax
  \def\url#1{\texttt{#1}}\fi
\expandafter\ifx\csname urlprefix\endcsname\relax\def\urlprefix{URL }\fi

\bibitem{BBCR10}
B.~Bollob\'as, C.~Borgs, J.~Chayes, and O.~Riordan, Percolation on dense graph
  sequences. \textit{Annals of Probability} \textbf{38}, 150--183 (2010).

\bibitem{KNZ14}
B.~Karrer, M.~E.~J. Newman, and L.~Zdeborov\'a, Percolation on sparse networks.
  \textit{Phys. Rev. Lett.} \textbf{113}, 208702 (2014).

\bibitem{Bonacich87}
P.~F. Bonacich, Power and centrality: A family of measures. \textit{Am. J.
  Sociol.} \textbf{92}, 1170--1182 (1987).

\bibitem{Fiedler73}
M.~Fiedler, Algebraic connectivity of graphs. \textit{Czech. Math. J.}
  \textbf{23}, 298--305 (1973).

\bibitem{PSL90}
A.~Pothen, H.~Simon, and K.-P. Liou, Partitioning sparse matrices with
  eigenvectors of graphs. \textit{SIAM J. Matrix Anal. Appl.} \textbf{11},
  430--452 (1990).

\bibitem{Newman06b}
M.~E.~J. Newman, Modularity and community structure in networks. \textit{Proc.
  Natl. Acad. Sci. USA} \textbf{103}, 8577--8582 (2006).

\bibitem{CM11}
M.~Cucuringu and M.~W. Mahoney, Localization on low-order eigenvectors of data
  matrices. Preprint arxiv:1109.1355 (2011).

\bibitem{MZN14}
T.~Martin, X.~Zhang, and M.~E.~J. Newman, Localization and centrality in
  networks. \textit{Phys. Rev. E} \textbf{90}, 052808 (2014).

\bibitem{PG16}
M.~A. Porter and J.~Gleeson, \textit{Dynamical Systems on Networks: A
  Tutorial}. Springer, Berlin (2016).

\bibitem{DKMZ11a}
A.~Decelle, F.~Krzakala, C.~Moore, and L.~Zdeborov\'a, Inference and phase
  transitions in the detection of modules in sparse networks. \textit{Phys.
  Rev. Lett.} \textbf{107}, 065701 (2011).

\bibitem{NN12}
R.~R. Nadakuditi and M.~E.~J. Newman, Graph spectra and the detectability of
  community structure in networks. \textit{Phys. Rev. Lett.} \textbf{108},
  188701 (2012).

\bibitem{DGMS03}
S.~N. Dorogovtsev, A.~V. Goltsev, J.~F.~F. Mendes, and A.~N. Samukhin, Spectra
  of complex networks. \textit{Phys. Rev. E} \textbf{68}, 046109 (2003).

\bibitem{RCKT08}
T.~Rogers, I.~P\'erez~Castillo, R.~K{\"u}hn, and K.~Takeda, Cavity approach to
  the spectral density of sparse symmetric random matrices. \textit{Phys. Rev.
  E} \textbf{78}, 031116 (2008).

\bibitem{RPTP10}
T.~Rogers, C.~P\'erez~Vicente, K.~Takeda, and I.~P\'erez~Castillo, Spectral
  density of random graphs with topological constraints. \textit{J. Phys. A}
  \textbf{43}, 195002 (2010).

\bibitem{MNB11}
F.~L. Metz, I.~Neri, and D.~Boll\'e, Spectra of sparse regular graphs with
  loops. \textit{Phys. Rev. E} \textbf{84}, 055101 (2011).

\bibitem{SC02b}
G.~Semerjian and L.~F. Cugliandolo, Sparse random matrices: The eigenvalue
  spectrum revisited. \textit{J. Phys. A} \textbf{35}, 4837--4852 (2002).

\bibitem{Kuhn08}
R.~K{\"u}hn, Spectra of sparse random matrices. \textit{J. Phys. A}
  \textbf{41}, 295002 (2008).

\bibitem{NZN19}
M.~E.~J. Newman, X.~Zhang, and R.~R. Nadakuditi, Spectra of random networks
  with arbitrary degrees. Preprint arxiv:1901.02029 (2019).

\bibitem{Newman09b}
M.~E.~J. Newman, Random graphs with clustering. \textit{Phys. Rev. Lett.}
  \textbf{103}, 058701 (2009).

\bibitem{Miller09}
J.~C. Miller, Percolation and epidemics in random clustered networks.
  \textit{Phys. Rev. E} \textbf{80}, 020901 (2009).

\bibitem{KN10b}
B.~Karrer and M.~E.~J. Newman, Random graphs containing arbitrary distributions
  of subgraphs. \textit{Phys. Rev. E} \textbf{82}, 066118 (2010).

\bibitem{EKBV05}
M.~Eckstein, M.~Kollar, K.~Byczuk, and D.~Vollhardt, Hopping on the {B}ethe
  lattice: Exact results for densities of states and dynamical mean-field
  theory. \textit{Phys. Rev. B} \textbf{71}, 235119 (2005).

\bibitem{BMN12}
D.~Boll\'e, F.~L. Metz, and I.~Neri, On the spectra of large sparse graphs with
  cycles. Preprint arxiv:1206.1512 (2012).

\bibitem{JGN01}
E.~M. Jin, M.~Girvan, and M.~E.~J. Newman, The structure of growing social
  networks. \textit{Phys. Rev. E} \textbf{64}, 046132 (2001).

\bibitem{HK02b}
P.~Holme and B.~J. Kim, Growing scale-free networks with tunable clustering.
  \textit{Phys. Rev. E} \textbf{65}, 026107 (2002).

\bibitem{KE02}
K.~Klemm and V.~M. Eguiluz, Highly clustered scale-free networks. \textit{Phys.
  Rev. E} \textbf{65}, 036123 (2002).

\bibitem{SB05}
M.~A. Serrano and M.~Bogu{\~n}\'a, Tuning clustering in random networks with
  arbitrary degree distributions. \textit{Phys. Rev. E} \textbf{72}, 036133
  (2005).

\bibitem{McKay81}
B.~D. McKay, The expected eigenvalue distribution of a large regular graph.
  \textit{Linear Algebra Appl.} \textbf{40}, 203--216 (1981).

\bibitem{MR95}
M.~Molloy and B.~Reed, A critical point for random graphs with a given degree
  sequence. \textit{Random Structures and Algorithms} \textbf{6}, 161--179
  (1995).

\bibitem{NSW01}
M.~E.~J. Newman, S.~H. Strogatz, and D.~J. Watts, Random graphs with arbitrary
  degree distributions and their applications. \textit{Phys. Rev. E}
  \textbf{64}, 026118 (2001).

\end{thebibliography}
\end{document}